\documentclass{jltp}
\usepackage{epsfig}

\title{A Single Impurity in a Luttinger Liquid: \\ How it ``Cuts'' the
  Chain\footnote{Dedicated to P.~W\"olfle on the occasion of his 60th 
birthday}}

\author{V.~Meden$^1$, W.~Metzner$^2$, U.~Schollw\"ock$^3$, 
and K.~Sch\"onhammer$^1$}

\address{$^1$Inst.\ f.\ Theoretische Physik, Universit\"at
  G\"ottingen, 37073 G\"ottingen, Germany\\
$^2$Inst.\ f.\ Theoretische Physik C, RWTH Aachen, 52056 Aachen,
Germany, \\
$^3$Sektion Physik, Universit\"at M\"unchen, 80333 M\"unchen, Germany}

\runninghead{V.~Meden, W.~Metzner, U.~Schollw\"ock, 
and K.~Sch\"onhammer}{A Single Impurity in a 
Luttinger Liquid}

\begin{document}

\maketitle

\begin{abstract}
Using a fermionic renormalization group method we present a simple
real space picture of the strong influence  an impurity has on the
electronic properties of a Luttinger liquid. We compute the flow of 
the renormalized impurity potential for a single impurity over the 
entire energy range - from the microscopic scale of a lattice-fermion 
model down to the low-energy limit. We confirm that low energy
properties close to the impurity are  
as if the chain is cut in two pieces with open boundary conditions 
at the end points, but show that this universal behavior 
is only reached for extremely large systems. The accuracy of the
renormalization group scheme is demonstrated by a direct comparison
with data obtained from the density-matrix renormalization group
method.  

PACS numbers: 71.10.Pm, 73.22.-f, 73.63.-b 
\end{abstract}

\section{INTRODUCTION}

The detailed theoretical understanding of the low-energy electronic 
properties of one-dimensional interacting electron systems obtained 
within the last fifty years has mainly been reached by 
mapping microscopic models onto an effective field theory using 
the very powerful technique of bosonization.\cite{Voitreview}
Bosonization, renormalization group (RG) investigations,\cite{Solyom} 
and the use of conformal field theory\cite{Voitreview} have led to
the concept of Luttinger liquid (LL) behavior which unifies the
low-energy physics of a wide class of models. 
In the mapping process terms which are expected to be irrelevant in the
low-energy limit are neglected and the energy scales of the
microscopic models get lost. Therefore a direct investigation of
lattice models is necessary to obtain informations about 
the {\it energy scales}\/ at which the 
typical LL power-laws and thus {\it universal behavior}\/ can be
expected. This is of special importance in connection with
the interpretation of experimental data. Such an investigation might 
furthermore lead to a more intuitive understanding of the
mechanisms involved in the physics of low-dimensional correlated
electrons. Here we present such a direct fermionic discussion
using analytical and numerical techniques focusing on the 
drastic influence a single impurity has on the low-energy physics of 
LL's.\cite{LutherPeschel,Mattis,ApelRice,KaneFisher,EggertAffleck,MatveevGlazman}

Within bosonization an impurity with scattering amplitudes $V_{k,k'}$,
($k$ denotes the wave vector) is modeled by forward and backward  
scattering only, i.e. $k$ and $k'$ are restricted to $\pm k_F$, where
$k_F$ is the Fermi wave vector.  
Both scattering channels are furthermore decoupled.  
A perturbative bosonic RG calculation\cite{KaneFisher} and a 
boundary conformal field theory analysis\cite{EggertAffleck} led to the 
following picture: For a chain of spinless fermions\cite{spinless} with 
repulsive interactions (LL parameter $K_{\rho}<1$) the backscattering
amplitude $V_B$ presents a relevant perturbation which grows as 
$\Lambda^{K_{\rho} - 1}$ when the flow parameter $\Lambda$ is sent 
to zero. This leads to a breakdown of the perturbative analysis. 
The scaling behavior can be traced back to the power-law 
singularity of the $2k_F$ density response function in a 
LL.\cite{LutherPeschel,Mattis} 
On the other hand\cite{KaneFisher} 
 a weak hopping $t_{w}$ between the open ends of 
two semi-infinite chains is irrelevant and scales to zero as 
$\Lambda^{K^{-1}_{\rho} - 1}$.
{\it Assuming} that the open chain presents the only stable fixed 
point it was argued that at low energy scales and even for a weak 
impurity physical observables behave as if the system was split in 
two semi-infinite chains with open boundary conditions at the end 
points.\cite{KaneFisher} 
Our main interest will be on the local spectral weight $\rho_j(\omega)$ 
for lattice sites $j$ close to the impurity and energies $\omega$ 
close to the chemical potential $\mu$. 
For $\rho_j(\omega)$ a power-law suppression 
$\rho_j(\omega) \sim |\omega|^{\alpha_B}$
with the {\it boundary exponent}\/ $\alpha_B = K_{\rho}^{-1} -1$ 
which only depends on the interaction strength and shape and 
filling of the band, but
{\it not} on the impurity parameters, was predicted.\cite{KaneFisher} 
Within the bosonic field theory the above 
assumption was verified by refermionization,\cite{KaneFisher} 
quantum Monte Carlo calculations,\cite{Moon,Reinhold} and the 
thermodynamic Bethe ansatz.\cite{Fendley} Although this gives a
consistent picture there are three important questions which have not 
been answered using field theory:
\begin{itemize}
\item In a more elaborate RG procedure the growing backscattering
  amplitude feeds back into the flow of all the other scattering
  channels $V_{k,k'}$ neglected in the field theory. Does this lead to other 
  stable fixed points? Does the above scenario hold 
  for microscopic lattice models?
\item Provided the scenario holds what is the scale on which 
  the universal open boundary fixed point (BFP) physics can be observed? 
\item Is it possible to obtain a simple physical picture of the 
  ``splitting'' mechanism? 
\end{itemize}

To numerically investigate the first two questions exact diagonalization 
(ED) and the density-matrix renormalization group (DMRG) method
were applied to the lattice model of spinless fermions with nearest neighbor 
interaction and a single impurity.\cite{EggertAffleck,Qin,Rommer,OC}   
The expected scaling was confirmed for both weak impurities and weak 
hopping using ED data.\cite{EggertAffleck}
However, due to the limited system size it was impossible to go 
beyond the perturbative (in either $V_B$ or $t_{w}$) regime. 
Later it was claimed that the full flow from a weak impurity to the
BFP was successfully demonstrated using DMRG,\cite{Qin,Rommer}
although this strong statement is not really supported by the 
numerical data presented. 
The smallest temperature discussed in Ref.\ \onlinecite{Rommer} corresponds 
to a system of around $300$ lattice sites and the largest system 
considered in Ref.\ \onlinecite{Qin} was $N=52$, 
while in Ref.\ \onlinecite{OC} it was shown that $N \approx 10^2$ 
lattice sites are clearly not enough to exclude an asymptotic
behavior not governed by the BFP, even if one starts out with a 
fairly strong impurity. 

An attempt to answer the above questions requires a method which is 
non-perturbative in the impurity strength, does give the LL power-law
divergence in the density response function, but
is not limited to systems of a few hundred lattice sites. 
We here use a functional RG method which has recently been introduced 
as a new powerful tool in the theory of interacting Fermi 
systems.\cite{Wetterich,Morris,Salmhofer1} Details of the derivation
of the flow equations are given and the RG scheme is applied to the 
spinless fermion model with site or hopping impurities.
In the extended numerical analysis of the flow equations (see
Sec.\ \ref{numsection}) we mainly focus on the hopping impurity.
This complements our earlier publication on the impurity 
problem.\cite{earlier} 
In our RG approach the complete flow of the 
renormalized on-site energies and the renormalized hopping amplitudes 
from the microscopic energy scale down to the infrared fixed point is
calculated. The flow equations are 
{\em non-perturbative in the impurity strength}\/ while perturbative 
in the electron-electron interaction.
We treat the {\em full functional form}\/ of the renormalized 
impurity potential as generated by the flow, instead of replacing it 
approximately by the scattering amplitudes at the Fermi level.
Computing the local spectral weight near the impurity we can 
confirm the flow to the BFP only if we start out with already fairly
large impurities even for systems of up to $10^4$-$10^5$ lattice
sites. This shows that for intermediate impurity and interaction 
parameters extremely large systems are required to reach the universal 
BFP. The quality of the approximations involved in our 
RG scheme is demonstrated by a direct comparison with essentially 
exact DMRG data for systems with up to $N=512$ sites. 

\section{THE MODEL}

The one-dimensional lattice model of spinless fermions with nearest
neighbor hopping amplitude $t=1$, lattice constant $a=1$, 
and nearest neighbor interaction $U$ is given by
\begin{eqnarray}
H = - \sum_{j} \left( c_j^{\dag} c_{j+1}^{} +
  c_{j+1}^{\dag} c_j^{}  \right)
  + U \sum_{j} n_j n_{j+1} ,
\label{spinlessfermdef}
\end{eqnarray}
in standard second-quantized notation. The boundary conditions we
consider are discussed below.
Here we mainly focus on the half filled band case for which the LL
parameter
\begin{eqnarray} 
K_{\rho}=\left[ \frac{2}{\pi} \arccos{ \left( - \frac{U}{2} \right)}
\right]^{-1} 
\label{Krho}
\end{eqnarray}
is analytically known from the Bethe 
ansatz.\cite{bethe} As already indicated by this
expression the model at half filling 
shows LL behavior only for $-2 < U < 2$. For $U>2$ the groundstate 
displays charge density wave order and for $U<-2$ phase separation 
sets in. We restrict ourselves to repulsive interactions in the LL 
regime, i.e. to $0< U<2$. 
To the Hamiltonian $H$ we either add a site impurity $H_s= V n_{j_0}$ 
or a hopping impurity $H_h = (1-t') (c_{j_0}^{\dag}
c_{j_0+1}^{} + \mbox{h.c.})$. 
$t'=0$ corresponds to a vanishing hopping between the sites $j_0$ and
$j_0+1$. Taking the thermodynamic limit or assuming open boundary
conditions in Eq.\ (\ref{spinlessfermdef}) a small $t'$ thus models 
the situation of a weak link between two decoupled chains. 
In the non-interacting limit the reflection
coefficient for scattering of such impurities 
at $k_F=\pi/2$ is given by\cite{OC} $|R_s|^2=V^2/(4+V^2)$ for $H_s$ and 
$|R_h|^2=(1-t'^2)^2/(1+t'^2)^2$ for $H_h$, 
which provides us with a measure for the strength of the bare impurity.
 
\section{A HARTREE-FOCK STUDY}

As a first step it is instructive to consider the impurity problem within
the Hartree-Fock approximation, before turning to the RG treatment.
The impurity leads to Friedel oscillations in the non-interacting 
density profile $\langle n_j \rangle_0$ which for large 
$|j-j_0|$ behaves as $|R| \sin{\left( 2k_F |j-j_0|\right)
  }/|j-j_0|$.\cite{phaseshift}  
Similar oscillations are found in the
matrix element $\langle c_j^{\dag} c_{j+1}^{} \rangle_0$.
Thus both the Hartree potential 
$V_j^H = U (\langle n_{j-1} \rangle_0 + \langle n_{j+1} \rangle_0 )$ 
and the Fock ``hopping correction''
$U ( \langle c_j^{\dag} c_{j+1}^{} \rangle_0 + \mbox{c.c.} )$
are oscillating and decay as $1/|j-j_0|$. 
One then has to solve a (non-trivial) one-particle problem within 
such a non-local potential. This problem is
interesting in itself - a special variant of it has been discussed 
in the mathematical literature.\cite{Klaus} 
Taking into account the Hartree term only,
the resulting spectral weight for $|\omega| \to 0$ shows power-law
behavior with an exponent which independently of the local part of the 
effective potential around $j_0$ is proportional to the amplitude
$U|R|$ of the 
asymptotic oscillations.\cite{Klaus}
It can be shown analytically that this behavior is not changed 
when the Fock term is included. 
A more detailed discussion of this scattering problem will be given
elsewhere.\cite{oscipaper} 
Thus solely due to the oscillations and the very slow $1/|j-j_0|$ 
decay of the non-local effective potential 
already HF yields a {\it power-law} for 
the spectral weight, but with an exponent which not only depends 
on $U$, but via $R$ also on the {\it bare}\/ impurity strength. 
This contradiction to the bosonization prediction is not surprising 
as HF does certainly not contain the renormalization of the impurity
strength. 

The extension of the HF study using self-consistent HF leads to
unphysical results and thus cannot be used to gain further insight. 
The self-consistent iterative solution of the HF equations generates 
for all $U>0$ a groundstate which shows charge density wave
order\cite{Richter} which is qualitatively incorrect since a single 
impurity cannot change bulk properties of the system.

\section{THE FERMIONIC RENORMALIZATION GROUP}

We now treat the problem using a fermionic functional RG
approach.\cite{Wetterich,Morris,Salmhofer1} Various versions were
applied recently to problems of strongly correlated two-dimensional
electron systems.\cite{2dsystems} Here we use the method proposed by 
Wetterich\cite{Wetterich} and Morris,\cite{Morris} where one
introduces a cut-off parameter $\Lambda$ in the free propagator $G^0$
cutting out degrees of freedom on energy scales less than $\Lambda$
and derives an exact hierarchy of coupled differential flow equations
for the one-particle irreducible vertex functions by differentiating
with respect to $\Lambda$, where $\Lambda$ flows from $\infty$ to $0$.

Before turning to the specific lattice model 
of spinless fermions with nearest neighbor interaction and hopping 
we will give a brief introduction to the
method for a general lattice model of spinless fermions
with interaction and impurity free
one-particle states $\left| \alpha \right>$ (e.g. Wannier states
$\left| j \right> $ or
momentum states $\left| k \right>$), 
two-body interaction
$\, \frac{1}{4} \sum_{\alpha,\beta,\gamma,\delta} 
\bar{v}_{\alpha,\beta,\gamma,\delta} \, 
 c^{\dag}_{\alpha} c^{\dag}_{\beta} c^{}_{\delta} c^{}_{\gamma}$, 
and an impurity potential 
$\, \sum_{\alpha,\beta} V_{\alpha,\beta} \, c^{\dag}_{\alpha} c^{}_{\beta}$.   
The {\it exact}\/ flow equation for the selfenergy reads
\begin{eqnarray}
 && \frac{d}{d \Lambda}\Sigma^{\Lambda}_{\alpha,\beta}(i \omega_n)  =  
 - T \sum_{\omega_l} e^{i\omega_l 0^+} \, 
 \sum_{\gamma,\delta} \Bigg\{   
 \left[  1- G^{0,\Lambda}(i \omega_l) 
 \Sigma^{\Lambda}(i \omega_l) \right]^{-1}  
 \frac{d G^{0,\Lambda}(i \omega_l)}{d \Lambda} \nonumber \\* 
 &&  \times \left[  1- \Sigma^{\Lambda}(i \omega_l)
    G^{0,\Lambda}(i \omega_l) \right]^{-1}
 \Bigg\}_{\delta,\gamma} \Gamma^{\Lambda}_{\alpha,\gamma,\beta,\delta}
 (i \omega_n,i \omega_l,i \omega_n,i \omega_l) ,
 \label{sflow}
\end{eqnarray}
where $G^{0,\Lambda}$ and $\Sigma^{\Lambda}$ are matrices with
matrix elements $G^{0,\Lambda}_{\alpha,\beta}$ and
$\Sigma^{\Lambda}_{\alpha,\beta}$, respectively,
$\omega_n$ denotes a fermionic Matsubara frequency, and 
$T$ is the temperature.
The irreducible four point vertex
$\Gamma^{\Lambda}_{\alpha,\gamma,\beta,\delta}$ 
obeys a flow equation with terms bilinear in $\Gamma^{\Lambda}$ 
and a term linear in the irreducible six point vertex. 
The initial condition is given by
$\Sigma^{\Lambda=\infty}_{\alpha,\beta} = V_{\alpha,\beta}$.
For spinless fermions the electron-electron interaction is
renormalized only by a finite amount of order interaction 
squared.\cite{Solyom} 
Hence as our central approximation we replace the renormalized 
two-particle vertex to leading order in the interaction by the 
antisymmetrized bare coupling, 
i.e.\ $\Gamma^{\Lambda}_{\alpha,\gamma,\beta,\delta} 
\to \bar{v}_{\alpha,\gamma,\beta,\delta}$. With this
approximation $\Sigma^{\Lambda}$ becomes {\it frequency independent.}
For the case of a model including a spin degree of freedom the flow of
the two-particle vertex has to be taken into account.\cite{Solyom}
In order to simplify the remaining Matsubara sum in Eq.\ (\ref{sflow})
we take the zero temperature limit and perform the RG flow with a {\it
  frequency cut-off} 
\begin{equation}
 G^{0,\Lambda}(i\omega) = \Theta(|\omega| - \Lambda) 
G^{0}(i\omega) .
\label{g0deff}
\end{equation}
The first factor on the right hand side (rhs) of Eq.\ (\ref{sflow}) then
involves products of a Dirac delta function and step functions. 
Using the relation\cite{Morris}
\begin{eqnarray}
\delta_{\varepsilon}(\omega-\Lambda) f\left( \Theta_{\varepsilon}
    \left[ \omega -
    \Lambda \right]\right) \to \delta(\omega - \Lambda) \int_{0}^{1}
   f(t) d \, t ,
\label{deltatheta}
\end{eqnarray}
where $\varepsilon$ is a broadening parameter tending to zero and $f$
is a sufficiently smooth function, we obtain
\begin{eqnarray}
 \frac{d}{d \Lambda}\Sigma^{\Lambda}_{\alpha,\beta} = 
 - \frac{1}{2 \pi}
 \sum_{\omega = \pm \Lambda} \sum_{\gamma,\delta}
 \bar{v}_{\alpha,\gamma,\beta,\delta} \, 
 G_{\delta,\gamma}^{\Lambda}(i \omega) \, e^{i \omega 0^+},
 \label{sigmalater}
\end{eqnarray}
where 
\begin{eqnarray}
G^{\Lambda}(i\omega) = \left\{ \left[ G^0( i \omega) \right]^{-1} -
  \Sigma^{\Lambda} \right\}^{-1}
\label{fullgdef}
\end{eqnarray}
is the full propagator for the cut-off dependent selfenergy.
The convergence factor $e^{i \omega 0^+}$ with $\omega = \pm\Lambda$
is relevant only for determining the flow from $\Lambda=\infty$ down
to some arbitrarily large finite $\Lambda_0$. 
For $\Lambda_0$ much larger than the band width this high energy 
part of the flow can be easily computed analytically and yields the
simple contribution 
$\frac{1}{2} \sum_{\gamma} \bar{v}_{\alpha,\gamma,\beta,\gamma}$. 
One can then drop the convergence factor and continue the flow
from $\Lambda_0$ downwards with the new initial condition
\begin{eqnarray}
 \Sigma^{\Lambda_0}_{\alpha,\beta} = V_{\alpha,\beta} + 
 {\textstyle\frac{1}{2}}
 \sum_{\gamma} \bar{v}_{\alpha,\gamma,\beta,\gamma} \; . 
\end{eqnarray}

Within this scheme $\Sigma^{\Lambda}_{\alpha,\beta}$ is the flowing effective
impurity potential. In our RG procedure all the different impurity 
scattering channels
corresponding to transitions from the one-particle states 
$\left| \alpha \right>$ to $\left| \beta \right>$ are coupled 
[see Eq.\ (\ref{sigmalater})] and we do not replace them by scattering
amplitudes at the Fermi level. This has to be contrasted to
the bosonization approach to the problem. 
The RG is a grand canonical method for which the chemical potential is
fixed and the average particle number has to be determined from the
Green function.  
To calculate observables, as
e.g. the local spectral weight, we have to determine the selfenergy 
$\Sigma^{\Lambda}_{\alpha,\beta}$  at $\Lambda=0$ and invert the 
matrix Eq.\ (\ref{fullgdef}) to
obtain the Green function, i.e. solve the one-particle problem of a
particle moving in the effective scattering potential
$\Sigma^{\Lambda=0}_{\alpha,\beta}$. This gives us a simple 
picture - in momentum or real space depending on the choice of 
one-particle states $\left| \alpha \right>$ - of the influence an 
impurity has on the electronic properties of a LL (see below). 

Because of the matrix inversion involved in calculating the rhs of 
Eq.\ (\ref{sigmalater}) we were not able to analytically 
solve the flow equation for general parameters. But it easy 
to show analytically that in the weak impurity limit 
Eq.\ (\ref{sigmalater}) exhibits the    
scaling predicted from bosonization. For that purpose we work in
momentum space and consider $\Sigma^{\Lambda}_{k,k'}$ for $k-k'$
different from a reciprocal lattice vector $K$. Then the term linear in
$\Sigma^{\Lambda}$ presents the leading approximation in the expansion
of $G^{\Lambda}$ on the rhs of Eq.\ (\ref{sigmalater}) and we obtain
\begin{eqnarray}
\frac{d}{d \Lambda}\Sigma^{\Lambda}_{k,k'} & = & - \frac{1}{2 \pi} \frac{1}{N}
\sum_{k_1,k_2} \sum_{K} \left[ \tilde{v}(k-k') - \tilde{v}(k-k_2) \right]
\delta_{k+k_1, k'+k_2+K} \nonumber \\*
&& \times \left[   \frac{1}{i \Lambda - \xi_{k_2} }
  \Sigma^{\Lambda}_{k_2,k_1} 
 \frac{1}{i \Lambda - \xi_{k_1} }  + \left( \Lambda \to - \Lambda \right) \right]
\label{expansion}
\end{eqnarray}
where $\xi_k = \varepsilon_k - \mu$ with the one-particle dispersion
$\varepsilon_k$, 
$\tilde{v}(k)$ is the Fourier transform of the two-particle
interaction, and $N$ is the number of lattice sites.
We treat the Hartree-type and Fock-type terms separately. The
Hartree-type term alone would lead to a selfenergy which depends on $k-k'$
only. In the thermodynamic limit one obtains
\begin{eqnarray}
\left[\frac{d}{d \Lambda}\Sigma^{\Lambda}_{k,k'}\right]_{H} & = & - 
\frac{1}{2 \pi} \tilde{v}(k-k') 
\int_{-\pi}^{\pi} \frac{d\,k_1}{2 \pi} \bigg[    \frac{1}{i \Lambda -
    \xi_{k-k'+k_1} }  \nonumber \\* && 
   \times  \Sigma^{\Lambda}_{k-k'+k_1,k_1}   \frac{1}{i \Lambda -
    \xi_{k_1} }  +  \left( \Lambda \to - \Lambda \right)  \bigg] .
\label{expansionH}
\end{eqnarray}
For $k=k_F$, $k'=-k_F$ the $k_1$-integration contains a singular
contribution proportional to $1/\Lambda$ which leads to the
power-law behavior with an exponent proportional to 
$\tilde{v}(2k_F)$ as discussed below. 
The Fock-type term reads 
\begin{eqnarray}
\left[\frac{d}{d \Lambda}\Sigma^{\Lambda}_{k,k'}\right]_{F} & = &  
\frac{1}{2 \pi} 
\int_{-\pi}^{\pi} \frac{d\,k_2}{2 \pi} \, \tilde{v}(k-k_2) 
 \bigg[    \frac{1}{i \Lambda -
    \xi_{k_2} }  \nonumber \\* && 
   \times  \Sigma^{\Lambda}_{k_2,k_2+k'-k}   \frac{1}{i \Lambda -
    \xi_{k_2+k'-k} }  +  \left( \Lambda \to - \Lambda \right)  \bigg] .
\label{expansionF}
\end{eqnarray}
Away from half filling the singular contribution $\sim 1/\Lambda$ for
$k=k_F$, $k'=-k_F$ comes from $k_2 \approx k_F$ and is therefore
proportional to $\tilde{v}(0)$. If only the singular contributions
are kept the differential equation for $\Sigma^{\Lambda}_{k_F,-k_F}$
reads 
\begin{eqnarray}
\frac{d}{d \Lambda}\Sigma^{\Lambda}_{k_F,-k_F} = - \left[
  \frac{\tilde{v}(0) -\tilde{v}(2 k_F) }{2 \pi v_F}  \right]
\frac{1}{\Lambda} \, \Sigma^{\Lambda}_{k_F,-k_F} ,
\label{sigmakfmkf}
\end{eqnarray}
where $v_F$ denotes the Fermi velocity. This leads to the scaling 
\begin{eqnarray}
\Sigma^{\Lambda}_{k_F,-k_F} \sim \left( \frac{1}{\Lambda} 
\right)^{[\tilde{v}(0) -\tilde{v}(2 k_F) ] /[2 \pi v_F]} .
\label{sigmascal}
\end{eqnarray}
To leading order in the interaction the exponent $[\tilde{v}(0)
-\tilde{v}(2 k_F) ] /[2 \pi v_F] $ is just
$1- K_{\rho}  $  which shows that our non-perturbative 
fermionic RG captures the power-law increase found in the 
perturbative bosonic RG. As Eq.\ (\ref{sigmakfmkf}) was derived by
expanding the Green function $G^{\Lambda}$ in powers of the
selfenergy, the scaling behavior  Eq.\ (\ref{sigmascal}) can be trusted
only as long as $\Sigma^{\Lambda}$ stays small. 
Eq.\ (\ref{sigmakfmkf}) also holds in the half filled band case since 
the two additional terms from the Hartree- and Fock-type contributions 
cancel each other. 
The scaling equation (\ref{sigmascal}) can similarly be
derived for a continuum model. 

\section{NUMERICAL SOLUTION OF THE RG EQUATIONS}

Numerically integrating the RG equations (\ref{sigmalater}) 
for finite systems of $N$ lattice sites we can go beyond the 
perturbative regime. In each step of the integration we have to 
invert an $N \times N$ matrix. 
If we now specialize on the lattice model of spinless fermions 
with nearest neighbor interaction and hopping, treat the problem 
in real space, and assume open boundary conditions, then  
$\left[  G^{0} (i \omega) \right]^{-1} - \Sigma^{\Lambda}  $ is
tridiagonal and Eq.\ (\ref{sigmalater}) reads
\begin{eqnarray}
\frac{d}{d \Lambda}\Sigma^{\Lambda}_{j,j} & = & - \frac{U}{2\pi} 
  \sum_{s=\pm 1} \sum_{\omega = \pm \Lambda}
  G^{\Lambda}_{j+s,j+s} (i \omega) 
  \label{diffsystem1} \\
\frac{d}{d \Lambda} \Sigma^{\Lambda}_{j,j\pm 1} & = & \frac{U}{2\pi} 
  \sum_{\omega = \pm \Lambda}
  G^{\Lambda}_{j,j \pm 1} (i \omega)  .
\label{diffsystem}
\end{eqnarray}  
Therefore  the numerical effort is considerably 
reduced which allowed us to treat systems with up to 
$2^{15} = 32768$ lattice sites. 
Since the position of the impurity $j_0$ is chosen to be far
away from the open boundaries they do  
not influence the behavior close to $j_0$. We have confirmed this 
by also investigating systems with periodic boundary conditions, in
which case we are limited to smaller systems of the order of  
$10^3$ lattice sites. The initial conditions for the case of a site 
impurity are 
$\Sigma^{\Lambda_0}_{1,1} = U/2 $, $\Sigma^{\Lambda_0}_{N,N} = U/2 $, 
$\Sigma^{\Lambda_0}_{j_0,j_0} = V+ U $, and
$\Sigma^{\Lambda_0}_{j,j} = U $
 for $j \neq j_0$ and not at the boundaries, while the
other matrix elements are initially zero. We started the integration
at $\Lambda_0=1000$ convincing ourselves that  further increasing 
$\Lambda_0$ only leads to negligible changes in the results.
For a hopping impurity we start with 
$\Sigma^{\Lambda_0}_{1,1} = U/2 $, $\Sigma^{\Lambda_0}_{N,N} = U/2$, 
$\Sigma^{\Lambda_0}_{j,j} = U $ 
and $\Sigma^{\Lambda_0}_{j_0,j_0+1} = 
 \Sigma^{\Lambda_0}_{j_0+1,j_0} = 1-t'$, while the
other matrix elements are again zero.

\begin{figure}[hbt]
\begin{center}
\epsfxsize7.0cm
\epsffile{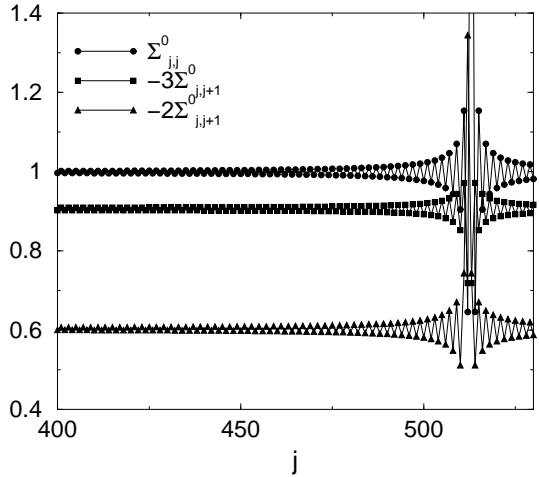}
\caption{$\Sigma_{j,j'}^{\Lambda=0}$ for a {\it site impurity}
  (circles and squares) with $V=1$ and a {\it hopping impurity} 
  (triangles) with $t'=0.5$, both for $U=1$. 
  For a clear representation the nearest neighbor parts are 
  multiplied by $-3$ respectively $-2$ as indicated in the
  legend. }  
\label{fig1}
\end{center}
\end{figure}
Fig.\ \ref{fig1} shows typical results for 
$\Sigma_{j,j}^{\Lambda=0}$ and $\Sigma_{j,j+1}^{\Lambda=0}$ 
for $U=1$, with a {\it site impurity} $V=1.5$ in one case 
and a {\it hopping impurity} $t'=0.5$ in the other,
where both would lead to a transmission probability of $|T|^2=0.64$ 
in a non-interacting system.  
Since $\Sigma^{\Lambda}$ is symmetric around $j_0$ and the ends of the
chain are dominated by the open boundaries mainly the region 
$1 \ll j<j_0$ is shown. For symmetry reasons in the 
first case we take $N=1025$, $j_0=513$ and the average
number of electrons at $\mu=U$ is $\left<N_F\right>=512$ so that we
are slightly off half filling. In the latter $N=1024$, $j_0=512$ 
and $\left<N_F\right>=512$. 
At half filling a hopping impurity leads to the shown 
long range oscillatory behavior in $\Sigma_{j,j+1}^{\Lambda=0}$
but {\it not} in  $\Sigma_{j,j}^{\Lambda=0}$, similar to the 
HF approximation.
 Furthermore
both the diagonal and the nearest neighbor part show global
shifts. For $\Sigma_{j,j}^{\Lambda=0}$ this shift by $U$ is due to
the initial condition and is exactly canceled by the chemical
potential which at half filling is given by $\mu=U$. A change in 
$\Sigma_{j,j+1}^{\Lambda=0}$ is present already in HF 
and corresponds to an interaction induced broadening of the band. 

\begin{figure}[h]
\begin{center}
\epsfxsize6.8cm
\epsffile{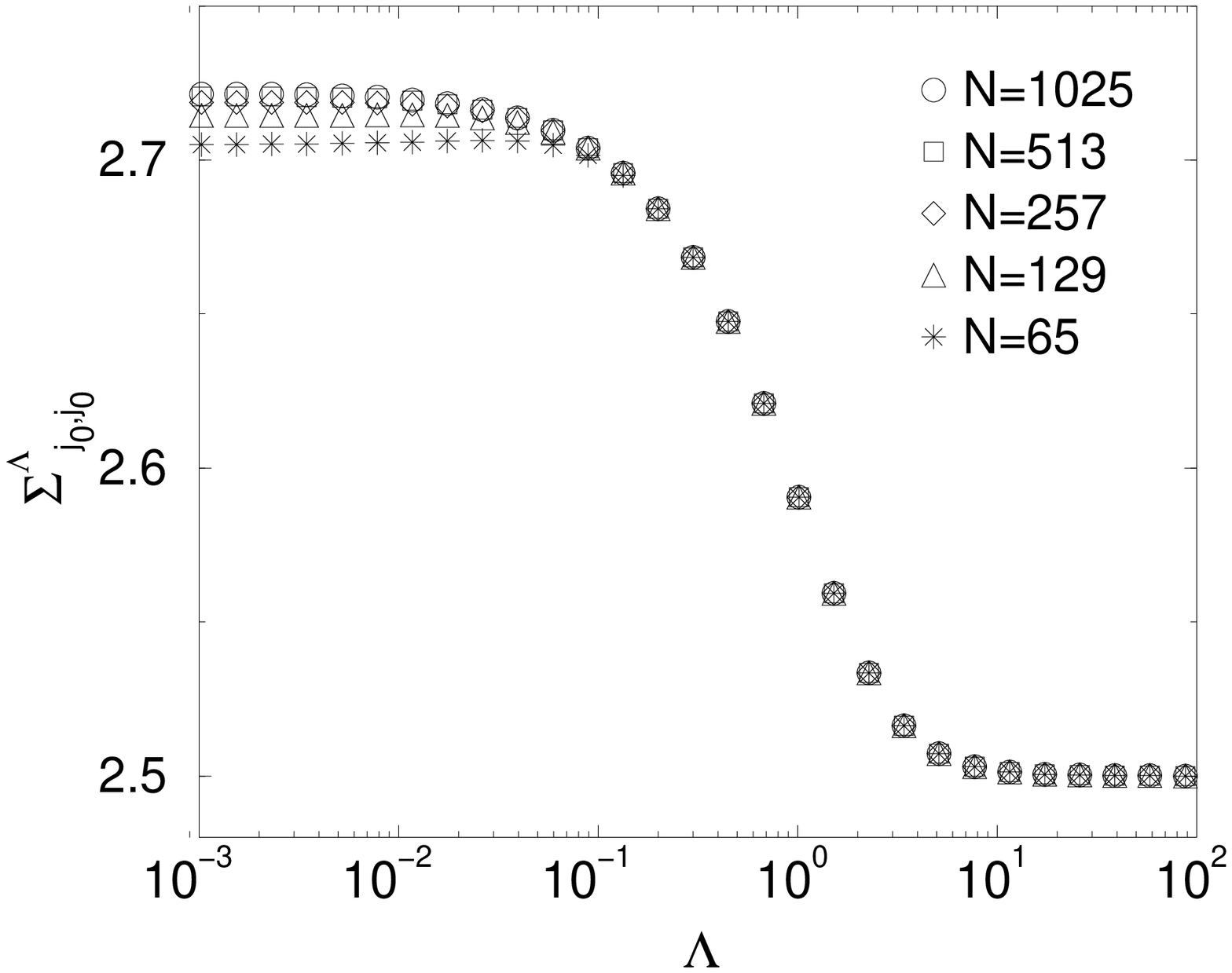}
\caption{$\Sigma_{j_0,j_0}^{\Lambda}$ as a function of $\Lambda$ for
  a {\it site impurity} (same parameters as in Fig.\ 1) 
  and different $N$.}  
\label{fig2}
\end{center}
\begin{center}
\epsfxsize6.8cm
\epsffile{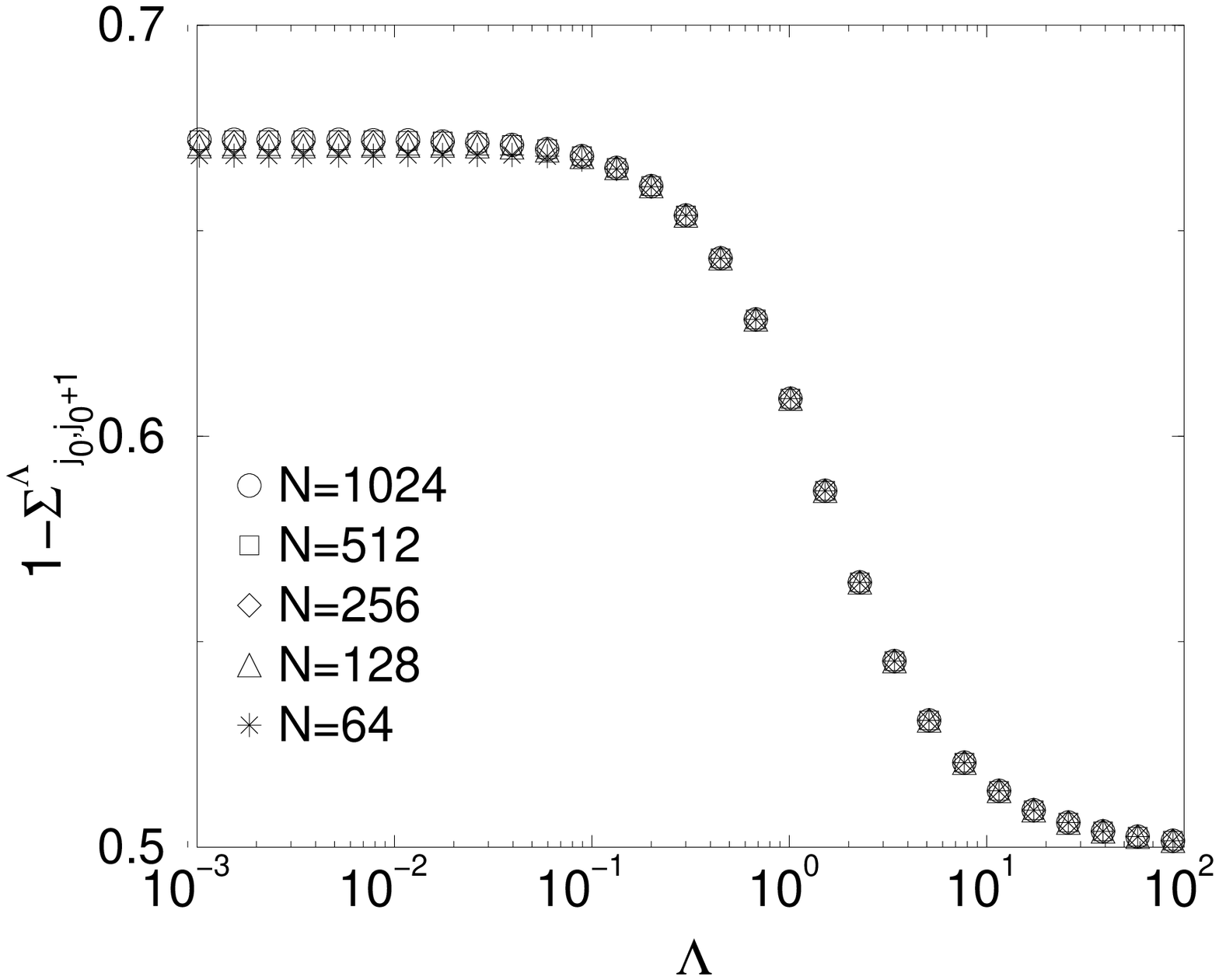}
\caption{The effective hopping $1-\Sigma_{j_0,j_0+1}^{\Lambda}$ 
between sites $j_0$ and $j_0+1$ as a function of $\Lambda$ for
  a {\it hopping impurity} (same parameters as in Fig.\ 1) 
  and different $N$.}  
\label{fig3}
\end{center}
\end{figure}

In Fig.\ \ref{fig2} we show 
$\Sigma_{j_0,j_0}^{\Lambda}$ for the {\it site impurity}
as a function of $\Lambda$ for different $N$ on a log-linear scale.
For finite $N$ the flow
is effectively cut off on a scale of the order of $1/N$ and to 
extrapolate to the thermodynamic limit we have to consider a 
sequence of $N$ values.
Obviously the renormalized potential at the impurity site remains 
finite in the limit $ N \to \infty$
and the expected ``cutting'' of the chain does certainly 
not occur because a single on-site energy diverges, as one might
guess if the bosonic RG is taken too literally. 
Singular behavior is only found in $\Sigma_{k,k'}^{\Lambda}$ 
for momenta with $k-k' \approx \pm 2k_F$, which is associated
with the {\it long range oscillations} in real space.

Fig.\ \ref{fig3} shows the effective hopping
$1-\Sigma_{j_0,j_0+1}^{\Lambda}$ 
for the {\it hopping impurity} between sites $j_0$ and $j_0+1$
as a function of $\Lambda$ for different $N$ on a log-linear scale.
 The effective hopping does 
{\it not tend to zero} for $N \to \infty$ as one could expect from a 
simplistic interpretation of the bosonization result.
As we do not perform a rescaling in our RG analysis the weak hopping 
has to stay finite because we otherwise would suppress the
transmission
through the weak link on {\it all} energy scales.
The
hopping between $j_0 $ and $j_0+1$ even increases upon 
integrating out degrees of freedom. 
Again the long range oscillations of the non-local effective impurity potential
and not the scaling 
of a single on-site energy or hopping matrix element 
is the reason for the peculiar behavior of physical observables, as 
for example $\rho_{j}(\omega)$, discussed next.

\begin{figure}[h]
\begin{center}
\epsfxsize7cm
\epsffile{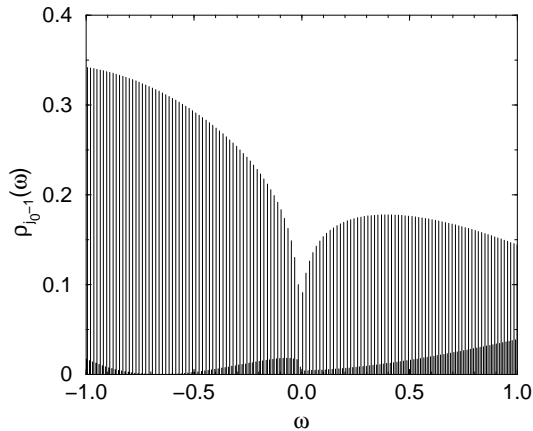}
\caption{$\rho_{j_0-1}(\omega)$ as a function of $\omega$
  for a {\it site impurity} with the same para\-me\-ters as in Fig.\ 1.}
\label{fig4}
\end{center}
\end{figure}
The local spectral function at the site $j_0-1$,
$\rho_{j_0-1}(\omega)$, is presented in Fig.\ \ref{fig4} 
for a {\it site impurity} with the same parameters as in Fig.\
\ref{fig1}.  
The  data show a suppression of the weight for $|\omega| \to 0$,
i.e. for energies close to the chemical potential,  
as expected from bosonization. A similar suppression is found in the
local spectral weight close to a hopping impurity.
Each spike represents a $\delta$-peak of the finite system.
For large systems the suppression follows a power-law
which similar 
to the HF description is related to the oscillatory
long range nonlocal effective potential. But while
 the HF exponent
depends on the bare impurity strength we expect our RG 
method  to lead  to a power law decay described by the boundary exponent 
$\alpha_B^{\rm RG}(U)$ {\it independent} of the the impurity
strength, where 
 $\alpha_B^{\rm RG}(U)$ can be determined from
the same RG analysis for an impurity free system
by calculating $\rho_j(\omega)$ for $j$ close to one of the
boundaries. 
The occurrence of $\alpha_B^{\rm RG}(U)$
can be shown analytically from Eqs.\ (\ref{diffsystem1})
 and (\ref{diffsystem}) only for $|V|\gg 1$ or $|t'|\ll 1$.  
We shortly sketch the argument for the weak hopping case.
If one introduces operators $\hat L$ and $\hat R$ which
project on the one-particle states to the left and right of the weak
link the resolvents $G^\Lambda(z)\equiv (z-h^\Lambda)^{-1}$ 
in Eqs.\ (\ref{diffsystem1})
 and (\ref{diffsystem}) can be expanded in the effective 
hopping matrix element $h^\Lambda_{j_0,j_0+1}
= -1+ \Sigma^\Lambda_{j_0,j_0+1}
\equiv  \tilde \Sigma^\Lambda_{j_0,j_0+1} $, anticipating that this quantity
is small for all values of $\Lambda$. One obtains
$G_{LL}(z)\equiv
\hat L G^\Lambda(z) \hat L=[z\hat L -\hat L h^\Lambda \hat
L]^{-1} +{\mathcal O} \left( \left[ \tilde \Sigma^\Lambda_{j_0,j_0+1}
  \right]^2 \right)$
 and a corresponding result for
$\hat R G^\Lambda(z) \hat R$. 
  If we denote the
leading term as $\left[G^\Lambda_{LL}(z)\right]_0$ (and  
$\left[G^\Lambda_{RR}(z)\right]_0$) 
and also expand 
$\hat L G^\Lambda(z) \hat R=\left[G_{LL}(z)\right]_0\hat L h^\Lambda 
\hat R G_{RR}(z)$
  the equation for 
 $\tilde \Sigma^\Lambda_{j_0,j_0+1}$ reads
\begin{equation}
\frac{d}{d\Lambda}\tilde \Sigma^\Lambda_{j_0,j_0+1}
=\left ( \frac{U'}{2\pi}\sum_{\omega=\pm \Lambda}
\left[G^\Lambda_{j_0,j_0}(i\omega)\right]_0
\left[G^\Lambda_{j_0+1,j_0+1}(i\omega)\right]_0 \right )
 \tilde \Sigma^\Lambda_{j_0,j_0+1}.
\label{weakhopping}
\end{equation}
 The correction terms to the rhs are of order
 $\left(\tilde \Sigma^\Lambda_{j_0,j_0+1} \right)^3$.
We also allowed the Coulomb integral $U'$ across the
weak link to be different from $U$. The equations for all other 
nonvanishing matrix elements of $\Sigma^\Lambda$, apart from higher
order corrections, are equations for a chain {\it split} between the
sites $j_0$ and $j_0+1$. Therefore it is obvious that local spectral
functions $\rho_j(\omega)$ for $j$ in the neighborhood of $j_0$
show a power law suppression with the boundary exponent 
 $\alpha_B^{\rm RG}(U)$. Once the equations
for the split chain are solved the result
for  $\tilde \Sigma^{\Lambda=0}_{j_0,j_0+1}$ follows from
Eq.\ (\ref{weakhopping}) by a simple integration. In the special case
$U'=0$ the effective hopping is given by $t'$ for {\it all}
 values of $\Lambda$.
The case of a strong site impurity $|V|\gg 1$ can be mapped to a weak
hopping problem with $t'\sim 1/V$ by projecting out the impurity site.

It is advantageous\cite{earlier}  
to analyze the finite size scaling of the 
spectral weight $W(N)$ at $\mu$ and $j_0-1$ 
instead of trying to fit a power-law
to the fixed $N$ data for $\rho_{j_0-1}(\omega)$ at small $\omega$.
If $\rho_{j_0-1}(\omega)$ follows a power-law as a function 
of frequency, we obtain a power-law with the same exponent in 
the $N$ dependence of $W(N)$.  
In the next section we investigate the behavior of the exponent
as a function of system size and
compare the data obtained within the RG scheme with essentially exact
data from DMRG focusing on hopping impurities.

\section{LARGE $N$ AND COMPARISON WITH DMRG RESULTS}
\label{numsection}

We have calculated the spectral weight $W(N)$ at $\mu$ and on the site
next to the modulated bond $j_0$ for different $t'$, $U$, and $N=2^n$.
Applying the RG scheme we were able to go up to $n=15$, while using
the DMRG we were limited to $n=9$. 
Since we are interested in the exponent of a possible  
power-law behavior we determine 
\begin{eqnarray}
\alpha_I(N=2^n) = - \frac{\ln{\left[ W\left(2^{n+1}\right)\right]}
- \ln{\left[ W\left(2^{n-1}\right)\right]} }{\ln{\left[ 2^{n+1}
  \right]} - \ln{\left[ 2^{n-1}
  \right]} } . 
\label{diffdeff}
\end{eqnarray}
If $W(N)$ decays for $N \to \infty$ as a power-law, 
$\alpha_I(N)$ converges to the respective exponent. 

\begin{figure}[h]
\begin{center}
\epsfxsize7cm
\epsffile{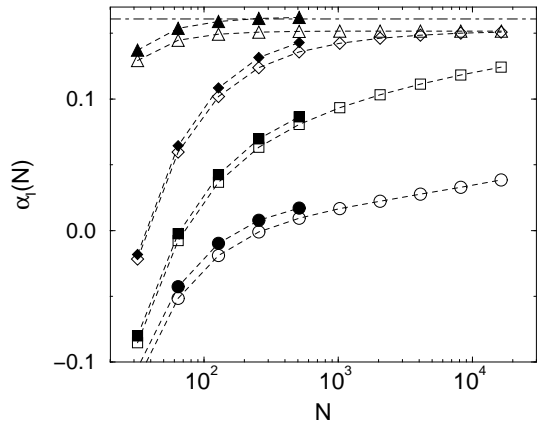}
\caption{$\alpha_I(N)$ as a function of $N$ for $U=0.5$ and different
  $t'$: $t'=0.8$ (circles), $t'=0.5$ (squares), $t'=0.2$ 
  (diamonds), and $t'=0$
  (triangles). The filled symbols are DMRG data, while the open ones
  have been obtained from the RG. 
  The dashed-dotted line gives the exact boundary
  exponent $\alpha_B^{\rm ex}$. }
\label{fig5}
\end{center}
\end{figure}

Fig.\ \ref{fig5} shows $\alpha_I(N)$ for $U=0.5$ and $t'=0.8$
(corresponding to $|T|^2 = 0.95$ at $U=0$, i.e. a very weak impurity), 
$t'=0.5$ ($|T|^2 = 0.64$ at $U=0$, i.e. an intermediate
impurity), and $t'=0.2$ ($|T|^2 = 0.15$ at $U=0$, 
i.e. a weak hopping). 
For comparison we also calculated $\alpha_B(N)$ for the lattice 
site next to an open boundary ($t'=0$).   
The DMRG and RG data are parallel to each other, which in addition
to the analytical arguments is a strong indication that our fermionic 
RG captures the essential physics.
For $t'=0$ both methods produce the expected power-law behavior 
with boundary exponents $\alpha_B^{\rm DMRG}$ and $\alpha_B^{\rm RG}$.
$\alpha_B^{\rm DMRG}(N=512)$ agrees up to $1$\% with the exact
exponent $\alpha_B^{\rm ex} \approx 0.1609$ obtained using Eq.\
(\ref{Krho}) and $\alpha_B^{\rm ex} = K_{\rho}^{-1}-1$.\cite{impnote} 
As our RG is only correct to leading order in $U$,   
 $\alpha_B^{\rm RG}(N=16384)$ which
effectively is equal to $\alpha_B^{\rm RG}(N=\infty)$
deviates by roughly $6$\% from $\alpha_B^{\rm ex}$. 
From the RG curves for small to intermediate bare hopping 
$t'$ between $j_0$ and $j_0+1$ one can infer that 
$\alpha_I^{\rm RG}(N)$ will approach the impurity independent 
boundary exponent $\alpha_B^{\rm RG}$ for $N \to \infty$, 
as predicted by bosonization. 
We have also determined $\alpha_I^{\rm RG}(N)$ for a site impurity
and three different values of $V$ leading to the same $U=0$ transmission 
probabilities $|T|^2$ as for the impurity parameters discussed 
here (see above). A comparison shows that for the hopping impurity 
and a fixed $|T|^2$ the convergence to $\alpha_B^{\rm RG}$ is slightly 
slower. Already for the site impurity extremely large 
$N = 10^4$-$10^5$ were needed to exclude non-universal 
(impurity dependent) fixed points with some certainty,\cite{earlier}
and here even longer chains are required. 

\begin{figure}[h]
\begin{center}
\epsfxsize7cm
\epsffile{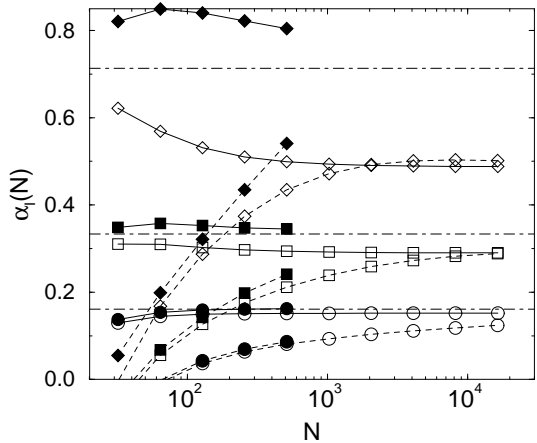}
\caption{$\alpha_I(N)$ as a function of $N$ for $t'=0.5$ (dashed lines)
  and $\alpha_B(N)$ for $t'=0$ (solid lines) for different 
  $U$: $U=0.5$ (circles), $U=1$
  (squares), and $U=1.8$ (diamonds). 
  Filled symbols are DMRG data, open ones RG results. 
  The dashed-dotted lines give the exact $U$ dependent 
  boundary exponents $\alpha_B^{\rm ex}$.}
\label{fig6}
\end{center}
\end{figure}

In Fig.\ \ref{fig6} RG and DMRG data are presented for an 
intermediate impurity strength $t'=0.5$ and $t'=0$ for 
different values of $U$. 
Due to higher order corrections in $U$, the
difference between the RG and DMRG data increases with increasing $U$.
For the case of a site impurity the speed of convergence of
$\alpha_I^{\rm RG}(N)$ to $\alpha_B^{\rm RG}$ was increased
considerably by increasing $U$ (see Fig.\ 4 
of Ref.\ \onlinecite{earlier}). 
In Fig.\ \ref{fig6} we also find a tendency towards a faster
convergence to $\alpha_B^{\rm RG}$ if $U$ is increased for a fixed $t'$, 
but the situation is less clear since the $\alpha_I^{\rm RG}(N)$ 
curves at large $N$ cross $\alpha_B^{\rm RG}(N)$ showing non-monotonic 
behavior and approach $\alpha_B^{\rm RG}$ from above. 

Compared to the case of a site impurity we here find somewhat 
less clear indications that the boundary exponent and thus the BFP 
is approached independently of the bare impurity strength. 
Nonetheless the presented RG data lead us to conclude
that the universality of the BFP holds also for a hopping impurity. 
 Our data show that for a hopping impurity and especially 
for intermediate $t'$ and $U$, which are experimentally most relevant, 
exceptionally large systems are needed to observe the universal BFP 
physics. For chains which are not long
enough a strong system size dependence  
of experimentally extracted exponents must be expected.
For both types of impurities it is obvious that a DMRG study alone, 
which is limited to a few hundred lattice sites, cannot give a definite 
answer to the question if the BFP is indeed universal.\cite{OC,earlier} 
To judge the quality of the approximations made to set up our RG
scheme the DMRG data are on the other hand very useful. 

\section{SUMMARY}

In summary, we have applied a functional RG method to a lattice
fermion model in order to analyze how and on what scale a single 
impurity in a Luttinger liquid ``cuts'' the chain. 
The method is perturbative in the electron-electron interaction
but non-perturbative in the impurity strength.
The results have been checked against essentially exact DMRG data for finite
systems, with good quantitative agreement for small $U$.
Computing the local spectral weight near a hopping impurity 
we found a clear indication of the expected BFP physics only for  
intermediate to strong hopping impurities even for systems of 
up to $N=2^{15}$ lattice sites. Having in mind the results of 
our previous study for site impurities\cite{earlier} we nevertheless 
conclude that universal power-law behavior with a
boundary exponent depending on the electron-electron 
interaction and not on the impurity strength also holds for hopping
impurities.
The functional RG method, where the entire functional form of the 
impurity potential is renormalized, provides a clear real space
picture of the cutting mechanism for both site and hopping impurities: 
while the local impurity strength is not renormalized much, an oscillating,
slowly decaying non-local potential develops which at low energies acts 
as an effective barrier.

It turned out that extremely large systems are required to reach 
the universal regime, even for intermediate impurity and 
interaction strengths. 
Hence only special mesoscopic systems, such as very long carbon
nanotubes, under very favorable experimental conditions are
suitable for experimentally observing the impurity induced
universal boundary physics. 
A trend {\it  towards}\/ cut chain behavior is certainly easier 
to observe. In Fig.\ \ref{fig4} e.g.\ a strong suppression of the local
spectral weight can be seen already for a chain with $N=1025$ lattice 
sites, a system size for which $\alpha_I(N)$ is still far away from 
the universal boundary exponent $\alpha_B$.  Functional RG methods 
applied to realistic models should be useful to compute 
the {\it non-universal}\/ behavior at intermediate scales, which 
is more easily accessible to experimentalists.

\section*{ACKNOWLEDGMENTS}
We would like to thank 
W.\ Apel, P.\ Durganandini, P.\ Kopietz, N.\ Shannon, C.\ Wetterich, 
and especially M.\ Salmhofer and H.\ Schoeller for valuable discussions. 
U.S.\ is grateful to the Deutsche Forschungsgemeinschaft for support
from the Gerhard-Hess-Preis.


\begin{thebibliography}{*}
\bibitem{spinless} Here we will mainly discuss spinless fermions.
\bibitem{phaseshift} We have neglected a phase shift which is
  irrelevant for the present discussion.
\bibitem{impnote} For $N=2^9$, $\alpha_{I/B}^{\rm DMRG}(N)$ was 
  determined by fitting $\ln\left[ W\left( 2^n \right)\right]$ 
  as a function of $\ln\left[ 2^n \right]$ by a spline and then taking
  the centered difference according to Eq.\ (\ref{diffdeff}).
\end{thebibliography}
\end{document}